\documentclass[aps,prb,twocolumn,reprint,showpacs]{revtex4-2}
\usepackage{bm}
\usepackage[dvips]{graphicx}
\usepackage[colorlinks, linkcolor = blue, citecolor = blue]{hyperref}
\usepackage[inline]{enumitem}

\usepackage{float}
\usepackage{graphics}

\usepackage{changes}
\usepackage{amsmath}
\usepackage{array}
\usepackage{makecell}

\newcommand{\unitdenom}{$(\Omega\times $cm$)^{-1}$}
\newcommand{\unit}{$(\hbar/e)(\Omega\times $cm$)^{-1}$}
\begin{document}

\title{Unconventional spin Hall effects in nonmagnetic solids}%

\author{Arunesh Roy, Marcos H. D. Guimar\~aes, and Jagoda S\l awi\'{n}ska}%
\affiliation{Zernike Institute for Advanced Materials, University of Groningen, Nijenborgh 4, 9747 AG Groningen, The Netherlands}
\begin{abstract}
Direct and inverse spin Hall effects lie at the heart of novel applications that utilize spins of electrons as information carriers, allowing generation of spin currents and detecting them via the electric voltage. In the standard arrangement, applied electric field induces transverse spin current with perpendicular spin polarization. Although conventional spin Hall effects are commonly used in spin-orbit torques or spin Hall magnetoresistance experiments, the possibilities to configure electronic devices according to specific needs are quite limited. Here, we investigate unconventional spin Hall effects that have the same origin as conventional ones, but manifest only in low-symmetry crystals where spin polarization, spin current and charge current are not enforced to be orthogonal. Based on the symmetry analysis for all 230 space groups, we have identified crystal structures that could exhibit unusual configurations of charge-to-spin conversion. The most relevant geometries have been explored in more detail; in particular, we have analyzed the collinear components yielding transverse charge and spin current with spin polarization parallel to one of them, as well as the longitudinal ones, where charge and spin currents are parallel. In addition, we have demonstrated that unconventional spin Hall effect can be induced by controllable breaking the crystal symmetries by an external electric field, which opens a perspective for external tuning of spin injection and detection by electric fields. The results have been confirmed by density functional theory calculations performed for various materials relevant for spintronics. We are convinced that our findings will stimulate further computational and experimental studies of unconventional spin Hall effects.

\end{abstract}
\maketitle

\section{Introduction}
In modern electronic devices that employ spin degrees of freedom, the mechanisms of conversion between spin and charge are essential to ensure all-electric control and low consumption of energy \cite{communications_materials}. The standard spin Hall effect (SHE) causes a transverse spin current (\textbf{$J_{S}$}) in response to a charge current (\textbf{$J_{C}$}) whose spin polarization (\textbf{$s$}) is perpendicular to both \textbf{$J_{S}$} and \textbf{$J_{C}$}. This effect is essential for many spintronics applications \cite{sinova_rev, jungwirth2012spin}, such as spin-orbit torques (SOT) which are widely exploited for magnetic memories \cite{chernyshov2009evidence, gambardella2011current, jabeur2014spin, prenat2015ultra, sot_memories}, and plays a crucial role in the spin Hall magnetoresistance effect \cite{huang2012transport, weiler2012local}. Because the intrinsic, and often dominant contribution to the SHE depends only on the electronic structure of a crystal, it has been intensely explored via first-principles calculations. Several materials with large spin Hall efficiencies have been revealed, most of them belonging to elemental metals with strong spin-orbit coupling (SOC) \cite{pt, tantalum, tungsten} as well as quantum materials with exotic band features, such as Weyl and Dirac nodal line semi-metals \cite{taas, felser, iro2}. Surprisingly, these efforts were mostly focused on finding materials with large magnitudes for the spin Hall conductivity (SHC) in conventional configurations, where the spin polarization, spin current and charge current are mutually orthogonal.
Nonetheless, the electrical generation of a spin current with out-of-plane spin polarization would be much more efficient for the manipulation of perpendicular magnetic anisotropy ferromagnets, such as the ones used in modern high-density memory devices. As this effect is forbidden in highly-symmetric conventional materials, the exploration of the spin Hall effects with unusual spin polarization and current directions is a promising field for future spintronic applications \cite{marcos_nature, manchon_2021}.

\begin{figure*}[htp]
    \includegraphics[width=0.99\textwidth]{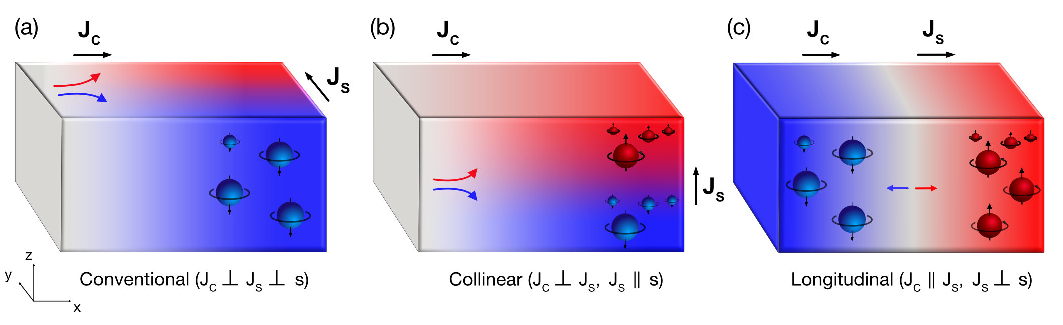}
    \caption{\label{fig1}
  Classification of the spin Hall effects. We choose the charge current along $x$ direction and illustrate (a) conventional SHE with spin current along $y$ and spin polarization along $z$ ($\sigma^{z}_{yx}$), (b) collinear SHE with spin current and spin polarization along $z$ ($\sigma^{z}_{zx}$), (c) longitudinal spin Hall effect with spin current along $x$ and spin polarization along $z$ ($\sigma^{z}_{xx}$).
    }
\end{figure*}

The spin Hall effect can be in general separated between extrinsic and intrinsic contributions \cite{sinova2015spin}. While the extrinsic component can originate from side-jump and skew-scattering by impurities, the intrinsic one is determined directly from the relativistic electronic structure; all these contributions can be calculated in linear response theory \cite{sinova2015spin, gradhand}. Because the charge and spin current are related through a third order spin Hall conductivity tensor whose indices may correspond to any spatial direction ($x, y$ or $z$), the charge-to-spin conversion could in principle occur in 27 different configurations. Surprisingly, the unconventional analogues of spin Hall effect (USHE) remained completely unexplored until few years ago, when they were theoretically predicted in non-magnetic crystals with low symmetry, \cite{anisotropy, wimmer2015spin, seemann2015symmetry} and ferromagnets where symmetry breaking is caused by the magnetization \cite{ferromagnets_perspective}. From the experimental side, only recently spin transport measurements in MoTe$_2$ revealed components with unusual spin polarization and spin Hall efficiency comparable with the conventional one \cite{vignale, casanova}. Although it demonstrated their huge potential for spintronics applications, few compounds allowing USHE have been so far proposed, and the efficient route to search or design materials with appropriate characteristics is still needed.

Here, we provide a symmetry-based analysis of the allowed SHC components - both conventional and unconventional - and use it to propose experimentally realizable systems for which the USHEs can be measured and/or controlled.
The symmetry analysis for the allowed components of the SHC tensors for all 230 space groups is performed using the tools implemented on the Bilbao Crystallographic Server (BCS) \cite{BCS1, BCS2}. By revealing the space groups for which unconventional charge-to-spin conversion configurations are allowed, we make possible the pre-selection of potential candidates for highly efficient manipulation of perpendicular magnetic anisotropy ferromagnets. In order to facilitate the discussion and further research, we have classified these spin Hall effects into the following categories; (i) conventional SHE, either isotropic or anisotropic, where \textbf{$J_{C}$}, \textbf{$J_{S}$}, and \textbf{$s$} are mutually orthogonal; (ii) collinear SHE, where the charge and spin currents are transverse, but spin polarization is parallel to one of them, i.e. \textbf{$J_{s}$} $\perp$ \textbf{$J_{C}$} and \textbf{$J_{S,C}$} $\|$ \textbf{$s$}; and (iii) longitudinal SHE where the charge current is parallel to the spin current (\textbf{$J_{S}$} $\|$ \textbf{$J_{C}$}) and the spin polarization is unconstrained. Based on these considerations, we have identified materials belonging to each class, calculated spin Hall conductivities using \textit{ab initio} methods, and whenever possible compared the results with the existing experimental data. Importantly, we have also revealed that unconventional spin Hall effect can be controlled by an external electric field, which has been confirmed by first principles calculations in a prototypical material (SnTe).

The paper is organized as follows: Section \ref{SC} gives a brief overview of spin Hall conductivity in the framework of linear response theory. In Section \ref{SAC}, we describe the employed methodology, namely symmetry analysis based on the TENSOR program \cite{gallego2019automatic} as well as details of the first principles simulations. In Section \ref{RandD}, we discuss the results of symmetry analysis complemented by numerical predictions for various materials relevant for spintronics. Section \ref{SnTe} discusses unconventional spin Hall conductivity induced by applied electric fields. In Section \ref{Conclusions}, we formulate the conclusions and suggestions for further computational and experimental studies. The Appendix summarizes allowed configurations for the spin Hall effects for all 230 crystallographic space groups.


\section{Intrinsic spin Hall conductivity} \label{SC}

In the framework of linear response theory, the spin current is expressed as:
\begin{equation}\label{spin_current}
J_{j}^{i} = \sigma_{jk}^{i} E_{k}\; \Rightarrow \bm{J}^{i} = \sigma^{i} \bm{E},
\end{equation}
which means that an applied electric field $E_{k}$ induces a spin current along $j$ with spin polarization along $i$. The current $J_{j}^{i}$ here is a 9-component tensor corresponding to different directions of spin current and its spin polarization \cite{anisotropy}. The spin current $\bm{J}$ is then related to the electric field via the spin Hall conductivity tensor:
\begin{equation}
    \sigma^{i} = \left(
    \begin{array}{ccc}
         \sigma_{xx}^{i}&\sigma_{xy}^{i}&\sigma_{xz}^{i}\\
         \sigma_{yx}^{i}&\sigma_{yy}^{i}&\sigma_{yz}^{i}\\
         \sigma_{zx}^{i}&\sigma_{zy}^{i}&\sigma_{zz}^{i}
    \end{array}
    \right),
\end{equation}
where $i = x,y,z$ are the spin polarization directions.

The Eq. (\ref{spin_current}) implies that the spin Hall conductivity $\sigma_{jk}^{i}$ is a third-order tensor with $3~(\mbox{electric field direction}) \times 3~ (\mbox{spin current direction}) \times 3~(\mbox{spin polarization}) = 27$ independent components since each index can be set along any of the spatial directions $x,y,z$.
This means that the expression for the spin Hall effect in linear response theory \cite{sinova2015spin}, can refer to different configurations of spin-to-charge conversion with essentially the same origin.

The form of the spin Hall conductivity tensor can be, in most cases, derived based on the Laue class of a crystal, as shown by Seemann \textit{et al.} \cite{seemann2015symmetry}. Nevertheless, it is more convenient to use the classification in terms of 230 space groups because the latter is typically used to describe a specific material \cite{glazer2012space}. We will thus consider the space groups for the rest of the paper for the sake of accessibility and include the other classifications in the tables in the Appendix. As we show in the next sections, the form of the SHC tensor is governed by the symmetries which may enforce several components to be zero.

In analogy to the anomalous Hall effect, the SHC tensor can be expressed by the Kubo formula \cite{sinova2015spin, gradhand}:
\begin{equation}\label{sigma_ijk}
\sigma_{jk}^{i} = -
\left(\frac{e}{\hbar}\right)
\int \frac{d^3 \bm{k}}{(2\pi)^3}
\sum_{n}f(\epsilon_{n,k}) \Omega^{i}_{jk,n}(\bm{k})
\end{equation}
in which $f(\epsilon_{n,k})$ is the Fermi-Dirac distribution function and the spin Berry curvature of the $n$th band is defined as:
\begin{equation}\label{berry_curvature}
\Omega^{i}_{jk,n}(\bm{k}) = \hbar^2 \sum_{m\neq n}
\frac{- 2\mbox{Im}\{\langle n\bm{k}|\mathcal{\hat{J}}^{i}_{j} |m\bm{k}\rangle \langle m\bm{k}|\hat{v}_{k} |n\bm{k}\rangle \}}{(\epsilon_{n,k}-\epsilon_{m,k})^2}
\end{equation}
where $\hat{v}_{i} = \frac{1}{\hbar} \partial \hat{H}/\partial k_i$ denotes the velocity operator and $\mathcal{ \hat{J}}^{i}_{j}=\{\hat{v}_{i},\hat{\sigma}_{j}\}/2 = (\hat{v}_{i}\hat{\sigma}_{j}+\hat{\sigma}_{j}\hat{v}_{i})/2$ is the spin current operator. Equation (\ref{berry_curvature}) quantitatively describes the deflection of the electron trajectories caused by spin–orbit interaction that can intrinsically occur in materials. It is governed by the strength of the spin-orbit coupling as well as the magnitude of the velocity and spin velocity vectors in a specific direction in momentum space. Equations (\ref{sigma_ijk}-\ref{berry_curvature}) can be evaluated via various approaches, such as Korringa–Kohn–Rostoker (KKR) method \cite{KKR} and tight-binding (TB) Hamiltonians derived from first principles calculations using either Wannier interpolation or the projections of wave functions on pseudo-atomic orbitals (PAO) \cite{wannier1, wannier2, agapito1, agapito2}; the latter, used in this work, is described in detail in Sec. \ref{SAC}.

\section{Symmetry analysis and computational details} \label{SAC}
We used the TENSOR program of Bilbao Crystallographic Server (BCS) \cite{gallego2019automatic} to find the allowed spin Hall conductivity components $\sigma_{ij}^{k}$ for all 230 crystallographic space groups.
To this aim, we have started with expressing the tensor in Jahn's notation in accordance with the BCS convention.
The SHC tensor is axial pertaining to the presence of the spin in the spin Berry curvature (see Eq. (\ref{berry_curvature})) and independent in all three indices. It can be written as e\{V\}\{V\}\{V\}, where e denotes axial and \{\} symbolizes allowed values of all three spatial indices for the designated vector V.
Then, we have directly used TENSOR to determine which $\sigma_{ij}^{k}$ components are allowed by symmetry.
As for most space groups these allowed terms are not independent, we have additionally established the symmetry-based dependencies.
Our analysis can therefore serve as a useful guide not only for experiments but also for computational studies, reducing the number of tensor elements that need to be calculated. 

Materials simulations were performed using the density functional theory (DFT), as implemented in the Quantum Espresso package \cite{qe1, qe2}.
In the calculations, we treated the ion-electron interaction with the fully relativistic projected-augmented wave (or norm-conserving) pseudopotentials from the pslibrary database \cite{pslibrary} and expanded the electron wave functions in a plane-wave basis, setting the cutoff to 80 (150) Ry.
The exchange and correlation interaction was accounted for via the generalized gradient approximation (GGA) parameterized by the Perdew, Burke, and Ernzerhof (PBE) functional \cite{pbe}.
The atomic coordinates of the structures were relaxed with the convergence criteria for energy and forces to $10^{-6}$ Ry and $10^{-4}$ Ry/bohr, respectively.
We performed the Brillouin zone (BZ) sampling at the DFT level following the Monkhorst-Pack scheme \cite{monkhorst-pack}, and converged the grid sizes for each material.

The intrinsic contributions to the spin Hall conductivity given by Eq. (\ref{sigma_ijk})-(\ref{berry_curvature}) were calculated in a post-processing step.
First, the tight-binding Hamiltonians were constructed from the projections of \textit{ab-initio} wavefunctions on atomic orbital bases following the implementation in the \textsc{PAOFLOW} code \cite{paoflow1, paoflow2}.
After interpolating the Hamiltonians on the ultra-dense $k$-points grids converged separately for every considered compound, the spin Berry curvatures were computed and integrated over the BZ using the adaptive smearing method \cite{adaptive}.
The external electric fields, when applicable, were added at the level of the TB Hamiltonians. The computational details for specific materials are given in the Supplementary Information.

\section{Results and discussion} \label{RandD}
The allowed spin Hall conductivity components found via symmetry analysis are summarized in the Appendix in the form of separate tables for each space group.
Before we discuss them in detail, let us make a few general remarks on the relationships between the space groups symmetries and the occurrence of specific components.
We have noticed that the space groups possessing fewer symmetry operations allow more independent components of the SHC tensor.
For example, the simplest space groups, SG 1 and 2, allow for all 27 components.
The further we progress towards space groups with more symmetry elements, one or more components get connected to each other via symmetry operations.
The high symmetry space groups - SG 207 to 230 are found to possess only one independent SHC tensor component.

The symmetry analysis indicates that all space groups permit the conventional SHE components, which implies the universality of the spin Hall effect. It means that any conducting material with sufficiently strong SOC will manifest charge-to-spin conversion. Its strength will depend on the details of the electronic structure of the specific material and will be the easiest to achieve in metals, where the Fermi surface contains states with large spin Berry curvature. Nevertheless, semiconductors may also exhibit SHE provided that their Fermi level lies within the valence or conduction band, as long as those also present a significant spin Berry curvature. The active control of the Fermi level and symmetries in semiconductors through (electrical) doping, strain or electric field, makes them particularly interesting for externally tuning the strength of SHEs.

While the universal existence of the SHE could be expected, its demonstration is not straightforward.
Here, we deduce it using symmetry considerations for any space group, following the arguments suggested by Mook \textit{et. al.} \cite{mook2020origin}.
Based on $S_{jk}^{i} = v_{i}\sigma_{k} v_{j}$, which is the symmetry equivalent quantity of the spin Berry curvature in Eq. (\ref{berry_curvature}), we determine the allowed SHC tensor components.
The quantity $S_{jk}^{i}$ is invariant under any symmetry operation of a particular space group.
Therefore, in the linear regime,
\begin{equation}
   \sigma_{jk}^{i} \equiv S_{jk}^{i} \xrightarrow[\mbox{Symmetry}]{\mbox{Operation}}
    \left\{
\begin{array}{cc}
     + S_{jk}^{i} & \mbox{allowed}  \\
    - S_{jk}^{i} & \mbox{not allowed}
\end{array}
    \right.
\end{equation}
This demonstrates that if $S_{jk}^{i}$ is positive under a chosen symmetry operation (e.g. translation, rotation, mirror reflection and inversion operation), the corresponding component of the spin Hall conductivity is allowed; otherwise, it will be prohibited by symmetry.

\subsection{Conventional spin Hall effects}
In the conventional SHE, the charge current is transverse to both spin current and spin polarization.
The corresponding SHC tensor components, $\sigma_{jk}^{i}$ when $i \neq j \neq k$, yields six possible configurations (see Eq. \ref{spin_current}) that can be connected by symmetries.
The most efficient spin Hall conductors are often isotropic, and identified among elemental metals, such as Pt (SG 225) or $\beta$-W (SG 223) \cite{pt, beta-tungsten}.
This fact was recently rationalized by Zhang \textit{et al.} who correlated the large SHC with the presence of many mirror planes in highly symmetric crystals \cite{koepernik}.
Nonetheless, SHC anisotropy could be explored in spin-logic devices, where, for example, SOTs vary in strength depending on the charge current direction.
Low-symmetry crystals can be completely anisotropic with six independent conventional components, while an increase in symmetry leads to a more isotropic behavior, as shown in Table \ref{tab:shc_components}.

Among the known compounds with anisotropic SHE, are bulk $T_d$-WTe$_2$ (SG 31), which has six independent conventional configurations with values ranging from -15 to -200 \unit\ \cite{zhou2019intrinsic, zhao2020unconventional}, as well as metallic rutile oxides, such as IrO$_2$ (SG 136) with three independent components and magnitudes between 10 and -250 \unit\ \cite{felser}. We have also found large spin Hall conductivity in a pyrite-type structure, PtBi$_2$ (SG 205), whose symmetries allow two independent SHE components. The values that we estimated from first principles calculations are $\sigma_{xy}^{z}$ = 975 and $\sigma_{xz}^{y}$ = -742 \unit, which yield anisotropy of over 30~\%. The spin Hall tensors provided in Appendix are thus helpful to rationalize and interpret the existing results as well as to design alternative spin Hall materials.

\begin{table}
\caption{\label{tab:shc_components} Conventional components by the number of independent components present in all space groups.}
\begin{ruledtabular}
\begin{tabular}{lllll}
Independent Components & 6 & 3 & 2 & 1\\
Space groups & 1-74 & 75-194 &195-206 &207-230\\
\end{tabular}
\end{ruledtabular}
\end{table}

\subsection{Collinear spin Hall effects}
\begin{figure*}
    \centering
    \includegraphics[width=0.99\textwidth]{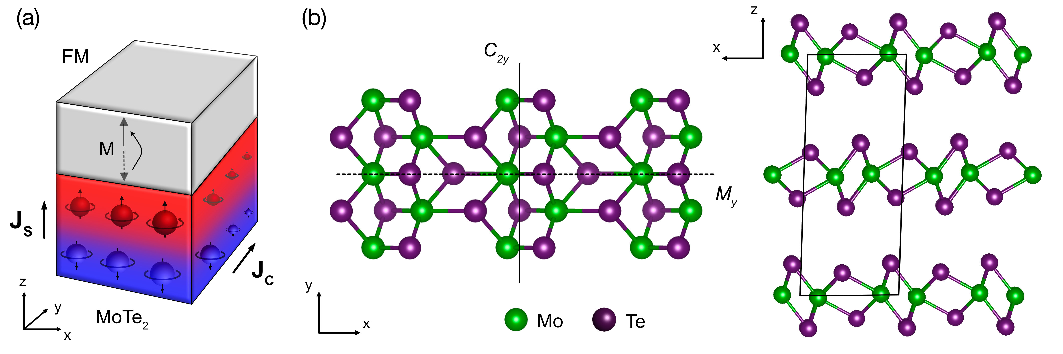}
    \caption{(a) Schematic view of the charge-to-spin conversion generating a spin-orbit torque in the bilayer consisting of MoTe$_2$ and the ferromagnet (FM). The charge current ($J_C$) induces a spin current ($J_S$) with a collinear spin polarization in a direction perpendicular to the planes $\sigma_{zy}^{z}$. The spin current then exerts a torque on the magnetization ($M$) of the ferromagnetic layer. (b) Top and side view of bulk 1T'-MoTe$_2$ crystal structure. The mirror plane ($M_y$) is denoted by a dashed line and the screw axis $C_{2y}$ is indicated by a solid line. The parallelogram defines the unit cell.}
    \label{mote2}
\end{figure*}

As we defined above, the collinear spin Hall conductivity is the configuration with transverse charge and spin current, and with the spin polarization aligned either with the charge current or with the spin current direction. The components of the SHC tensor are denoted as $\sigma_{jk}^{i}$ when $j\neq k$ and $i$ is equal to $j$ or $k$, e.g. $\sigma_{zx}^{z}$.
In Table \ref{sigma_shc_nconv}, we summarize the space groups that allow for collinear spin Hall effects.
Note that $\sigma_{ji}^{i}$ and $\sigma_{ij}^{i}$ always occur together, although in general they are not equal.
While the components with spin polarization parallel to the charge current were the first ones reported experimentally \cite{vignale, casanova}, those with spin currents parallel to spin polarization could be even more relevant for applications, as they can directly contribute to the generation of out-of-plane spin-orbit torques, schematically illustrated in Fig. \ref{mote2} (a).
The materials that seem to reveal the largest potential for unconventional SOTs are low-symmetry transition metal dichalcogenides (TMDs); we will thus analyze them in detail.

Let us first consider a bulk MoTe$_2$ which typically crystallizes in a semimetallic monoclinic phase (1$T'$ or $\beta$-MoTe$_2$, SG 11) with an inversion symmetry, two-fold screw axis along the $y$ direction (C$_{2y}$) and a mirror plane ($M_y$), as indicated in Fig. \ref{mote2} (b). Experiments typically employ slabs consisting of several layers. Odd-layered samples still belong to SG 11, while in the case of even number of layers, the crystal will be described by SG 6 \cite{raman}. We thus note that the bulk analysis remains valid, and the form of the SHC tensor will not change, preserving the same Laue class and the same non-vanishing components of SHC.
Our DFT calculations confirmed the presence of two pairs of collinear components, namely $\sigma_{xy}^{x}$ and $\sigma_{yx}^{x}$ as well as $\sigma_{zy}^{z}$ and $\sigma_{yz}^{z}$.
Even though their magnitudes are not large, the related spin Hall efficiencies (defined as ratios between spin Hall and charge conductivity, $\theta_{ij}^k = \sigma_{ij}^k / \sigma_c$, listed in Table III) may exceed 1 \%, mostly due to the rather low charge conductivity ($\sigma_c \sim 1800$ \unitdenom) of the semimetal \cite{hughes}.

These results are in a qualitative agreement with the measurements of $\sigma_{yz}^{z}$ in MoTe$_2$ \cite{vignale, hoque2021all}; nevertheless, the observed values of spin Hall angles are larger than predicted.
The possible reason is that extrinsic contributions, which are not included in the calculations, may lead to an additional increase in spin current.
Spin accumulation due to the Rashba-Edelstein effect could also play a role but additional studies will be needed in order to verify and numerically estimate its contribution. On the other hand, we can also compare the results with the measurements of unconventional spin-orbit torques generated by 1$T'$-MoTe$_2$ crystals \cite{stiehl}. The out-of-plane SOTs were observed for charge currents perpendicular to the mirror plane ($M_y$, see Fig. \ref{mote2} (b)), which in our nomenclature corresponds to the $\sigma_{zy}^{z}$ component. In contrast, the electric current flowing along the mirror plane did not yield any unconventional spin-orbit torque, which is again consistent with the theory, as the corresponding $\sigma_{zx}^{z}$ conductivity is forced to zero by symmetry. We emphasize that here the estimated spin Hall efficiencies agree quite well with the experiment, suggesting that USHE would play a major role in generating SOTs.

Similar analysis can be repeated for other TMDs. The experiments performed for WTe$_2$ revealed unconventional SOTs occurring in the same configurations as in 1$T'$-MoTe$_2$ \cite{macneill, macneill2017control}.
Although bulk WTe$_2$ crystallizes in an orthorhombic phase, described by SG 31, which yields only six independent conventional components, a few-layer system reduces to SG 6.
The symmetry of the SHC tensor will be thus the same as in the case of 1$T'$-MoTe$_2$ which explains the very similar experimental results.
In contrast, recent measurements performed for hexagonal TMDs, such as WSe$_2$ and MoS$_2$, have not shown any unconventional spin-orbit torques \cite{jan, shao2016strong, zhang2016research, novakov2021interface}. This can be again rationalized via a careful analysis of symmetries. The bulk SG 194 reduces to either SG 164 or SG 187 for respectively even and odd number of layers in the slab \cite{phonons}. As can be found in Appendix, these space groups do not allow for collinear spin Hall effects.

\begin{table}
\caption{\label{sigma_shc_nconv}
Allowed space groups for collinear SHE.
}
\begin{tabular}{|c|c|c|}
\hline
\hline
\thead{Components} & \thead{Allowed} & \thead{Not allowed} \\
\hline
$\sigma_{xy}^{x}$, $\sigma_{yx}^{x}$ &
\makecell{1,2, 3-15, 143-149,\\
151,153, 157, 159,\\
162-163} &
\makecell{16-142, 150, 152,\\
154-156, 158, \\
160-161, 164-230}
\\
\hline
$\sigma_{zx}^{x}$, $\sigma_{xz}^{x}$
&
\makecell{1,2, 75-88,\\
143-148, 168-176}
&
\makecell{3-74, 89-142,\\
149-167, 177-230}
\\
\hline
$\sigma_{xy}^{y}$, $\sigma_{yx}^{y}$
&
\makecell{1, 2, 143-148, 150,\\
152, 154-156, 158,\\
160–161, 164–167}
&
\makecell{3-142,  149, 151,\\
153, 157, 159, \\
162-163, 168-230}
\\
\hline
$\sigma_{yz}^{y}$, $\sigma_{zy}^{y}$
&
\makecell{1, 2, 75 -88, \\
143-148, 168–176}
&
\makecell{3-74, 89-142, \\
149–167, 177-230}
\\
\hline
$\sigma_{yz}^{z}$, $\sigma_{zy}^{z}$
&
\makecell{ 1,2, 3-15}
&
\makecell{16-230}
\\
\hline
$\sigma_{zx}^{z}$, $\sigma_{xz}^{z}$
&
\makecell{1,2}
&
\makecell{3, 4-230}
\\
\hline
\end{tabular}
\end{table}

\begin{table}[h]
    \label{mote2_shca}
    \caption{Spin Hall efficiencies $\theta_{ij}^k = \sigma_{ij}^k$ / $\sigma_c$ corresponding to independent SHC components calculated for 1T'-MoTe$_2$. We use the experimental value of $\sigma_c$ = 1.8$\times$ 10$^{3}$ \unitdenom \cite{hughes}. All spin Hall angles are expressed in \%.\\
    }
    \centering
    \begin{tabular}{|c|c|c|c|c|c|c|c|c|c|c|c|c|}
        \hline
        $\theta_{xy}^x$ & $\theta_{yy}^y$ & $\theta_{yx}^x$ &  $\theta_{zx}^y$ & $\theta_{yz}^x$ & $\theta_{zz}^y$ &  $\theta_{zy}^x$ & $\theta_{xy}^z$ & $\theta_{xx}^y$ & $\theta_{yx}^z$ &  $\theta_{xz}^y$ & $\theta_{yz}^z$ & $\theta_{zy}^z$\\ \hline
        -1.0 & 0.4 & -0.1 & 2.2 & 0.4 & 0.0 & 2.0 &
        -7.2 & -0.3 & 10.0 & -3.2 & 0.4 & 1.14 \\
        \hline
    \end{tabular}
\end{table}

\subsection{Longitudinal spin Hall effects}
In the longitudinal spin Hall effect, the applied electric field and induced spin current are parallel to each other whereas the spin polarization is independent. These SHC components are denoted as $\sigma_{jk}^{i}$ with $j=k$, and can be further classified in two categories. The first one will correspond to components with $i \neq j$, e.g. $\sigma_{xx}^{y}$, which are allowed in space groups listed in Table \ref{tab:sigma_ii_k}. The second one will refer to the special configurations in which $i = j = k$. They will describe USHE that are simultaneously longitudinal and collinear, representing a peculiar setup where electrons with spins parallel to the momentum are transmitted through a material and those with spins aligned anti-parallel are reflected, or vice versa. The corresponding space groups are summarized in Table \ref{tab:sigma_ii_i}.

Longitudinal spin currents have been hardly studied in non-magnetic systems.
1T'-MoTe$_2$ possesses three longitudinal components $\sigma_{xx}^{y}$, $\sigma_{zz}^{y}$ and $\sigma_{yy}^{y}$, but they have not been reported in experimental studies so far.
This is most likely due to the small values for the SHC (see Table III).
A recent computational high-throughout study revealed that a sizable $\sigma_{zz}^{z}$ component could be found in a metallic P$_7$Ru$_{12}$Sc$_2$ (SG 174), but again these findings need to be confirmed by experiments \cite{koepernik}.
Moreover, another simultaneously collinear and longitudinal component $\sigma_{xx}^{x}$ was predicted in a ferroelectric GeTe (SG 160) \cite{haihang}.
Conventional inverse spin Hall effect was detected in the heavily doped ferroelectric samples yielding $\theta_{\mathrm{SH}} \approx 1$ \% \cite{nat_electronics}, but the unusual component has not yet been measured; we expect its spin Hall efficiency to be lower than the conventional one. Further systematic search among semi-metals and (doped) semiconductors is needed in order to identify other potential candidate materials.


\begin{table}
\caption{\label{tab:sigma_ii_k} Allowed space groups for longitudinal SHE.}
\begin{tabular}{|c|c|c|}
\hline \hline
\thead{Components} & \thead{Allowed} & \thead{Not allowed} \\
\hline
$\sigma_{xx}^{y}$&
\makecell{1, 2, 3-15, 143-149,\\151, 153, 157, 159,\\162-163}&
\makecell{16-142, 150, 152,\\ 154-156, 158, 160\\-161, 164-230}\\
\hline
$\sigma_{xx}^{z}$, $\sigma_{yy}^z$ &
\makecell{1, 2, 75-88, \\ 143-148, 168-176}&
\makecell{3-74, 89-142, \\ 149-167, 177-230}\\
\hline
$\sigma_{yy}^{x}$ &
\makecell{1, 2, 143-148,\\ 150, 152, 154-156,\\ 158, 160-161, 164-\\167} &
\makecell{3-142, 149, 151,\\ 153, 157, 159, 162-\\163, 168-230}\\
\hline
$\sigma_{zz}^{x}$ &
\makecell{1,2}&
\makecell{3-230}\\
\hline
$\sigma_{zz}^{y}$ &
\makecell{1, 2, 3-15}&
\makecell{16-230}\\
\hline
\end{tabular}
\end{table}

\begin{table}
\caption{\label{tab:sigma_ii_i} Allowed space groups with SHE that is collinear and longitudinal at the same time.}
\begin{tabular}{|c|c|c|}
\hline \hline
\thead{Components} & \thead{Allowed} & \thead{Not allowed} \\
\hline
$\sigma_{xx}^{x}$&
\makecell{1 - 2, 143 - 148,\\
150, 152, 154 - 156,\\ 158, 160, 161, 164-167} &
\makecell{3 - 142, 149, 151, \\
153, 157, 159,\\
162, 168 - 230}\\
\hline
$\sigma_{yy}^{y}$ &
\makecell{1, 2, 3 - 15, \\
143-149, 151, 153, \\
157, 159, 162 - 163}&
\makecell{16 - 142, 150, 152,\\
154 - 156, 158, \\
160 - 161, 164 - 230}\\
\hline
$\sigma_{zz}^{z}$ &
\makecell{1 - 2, 75 - 88, \\143 - 148, 168 - 176} &
\makecell{3 - 74, 89 - 142,\\ 149 - 165, 177 - 230}\\
\hline
\end{tabular}
\end{table}

\section{Unconventional spin Hall effect induced by electric field} \label{SnTe}
\begin{figure*}[htp]
    \includegraphics[width=0.99\textwidth]{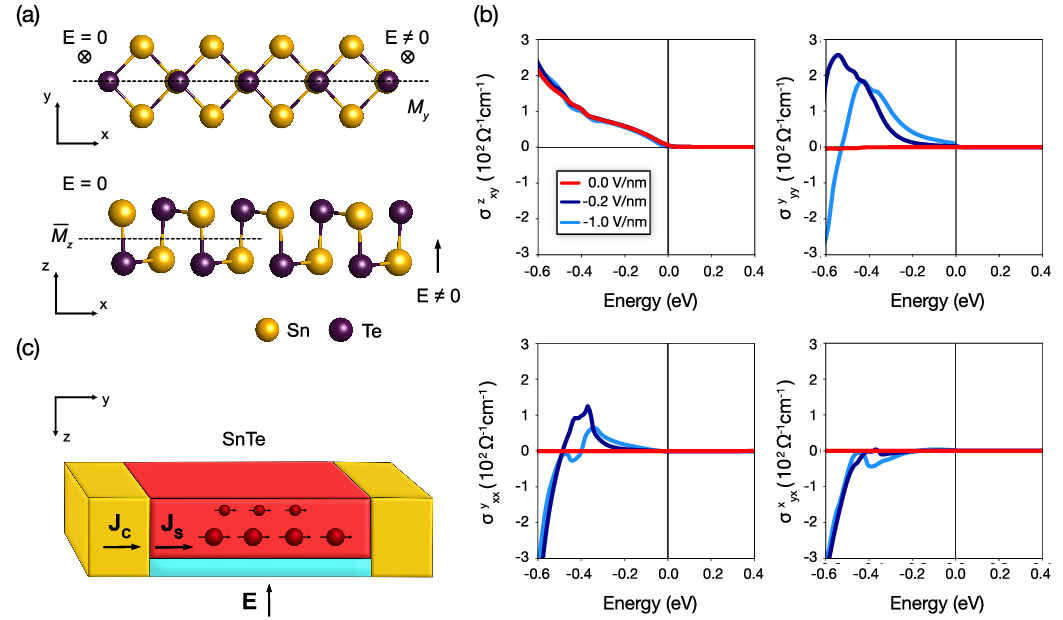}
    \caption{\label{snte}
  Electric field induced unconventional spin Hall effects in 2D-SnTe. (a) Top and side view of SnTe crystal structure. In the upper panel, the dashed line denotes the mirror plane ($M_y$) which remains the same in the presence of an out-of-plane electric field. In the bottom panel, the dashed line corresponds to the glide plane ($\overline{M}_{z}$). Together with the two-fold screw rotation ($\overline{C}_{2x}$) the glide plane vanishes upon the electric field, as shown in the right-hand side of the structure. (b) Spin Hall conductivity vs chemical potential with and without electric field. Because $\sigma^{z}_{yx}$ and $\sigma^{x}_{xy}$ are similar to respectively $\sigma^{z}_{yx}$ and $\sigma^{x}_{yx}$, the former have been omitted in the figure. The values of SHC are expressed in bulk units normalized using the effective thickness of 1ML-SnTe (approximately 10 \AA) \cite{ultrathin}. (c) Scheme of a prototypical device realizing spin injection induced by constant electric field perpendicular to 2D-SnTe.
    }
\end{figure*}

Finally, we will explore the possibility of inducing unconventional spin Hall effect by modifying crystal symmetry.
Such a control can be achieved \textit{in situ} using an external electric field or strain applied to any material provided that the space groups of the crystal without and with the stimulus are properly adjusted.
Here, we will consider the two-dimensional semiconductor SnTe, in its monolayer form (1ML-SnTe) consisting of two atomic layers as illustrated in Fig. \ref{snte} (a), which has been explored in the context of the conventional spin Hall effect \cite{2Dmaterials_perturbation, ultrathin}. The two-dimensional SnTe (2D-SnTe) is described by SG 31 and it is invariant under four symmetry operations: (i) identity $E$, (ii) mirror reflection ($M_y$) with respect to the $xz$ plane, (iii) glide reflection ($\overline{M}_{z}$) combining a mirror reflection with respect to the $xy$ plane and a fractional translation by a vector $\tau = (0.5a, 0.5b, 0)$ where $a$ and $b$ are the lattice constants, and (iv) two-fold screw rotation consisting of two-fold rotation around $x$ and the fractional translation by the vector $\tau$.
It can be seen that an electric field applied perpendicular to the plane (along $z$) will lift both the glide reflection and two-fold screw rotation, reducing the crystal structure from SG 31 to SG 6.
In accordance with the tensor forms listed in Appendix, the number of allowed spin Hall components would increase from six to thirteen.

Figure \ref{snte} (b) shows the spin Hall conductivities calculated for 2D-SnTe in the presence of an out-of-plane electric field $\vec{E}$.
First, we note that in a two-dimensional system neither charge nor spin transport can occur along the perpendicular direction ($z$) due to the lack of electronic dispersion with respect to $k_z$.
This eliminates all conventional and unconventional SHC tensor elements that describe configurations with out-of-plane currents; these zero SHCs are thus not shown.
Because the components $\sigma^{z}_{yx}$ and $\sigma^{x}_{xy}$ are similar to $\sigma^{z}_{yx}$ and $\sigma^{x}_{yx}$, respectively, they have also been omitted in Fig. \ref{snte} (b).
The most important result emerging from the plots is that $\sigma_{yy}^{y}$, $\sigma_{xx}^{y}$, $\sigma_{yx}^{x}$, $\sigma_{xy}^{x}$ are indeed induced by the electric field, which confirms our hypothesis derived from the symmetry analysis.
Notably, the component $\sigma_{yy}^{y}$ can achieve large magnitudes, comparable with the conventional spin Hall conductivity for energies close to $E_F$.

We also observe that the values of the induced SHCs depend on the magnitude of the electric field. This becomes clear by comparing the dark and light blue curves corresponding to fields of 0.2 and 1.0 V/nm, respectively.
The intrinsic spin Hall conductivity is entirely determined by the electronic structure, which is significantly altered by the electric field, therefore leading to a change in the SHC values over the entire energy range.
Although the conventional components seem to remain robust, we have found that for the opposite sign of $\vec{E}$, SnTe may become metallic, which means that spin Hall conductivity must be carefully estimated in each case.

Last, we note that unconventional spin Hall effect induced by an electric field could be detected using ferromagnetic voltage electrodes which probe locally the spin chemical potential in a device similar to the one presented in Fig \ref{snte} (c). Since the realization of experiments using a two-dimensional material seems more challenging, we emphasize that 1ML-SnTe could be replaced by a multilayer with the AA stacking. This configuration will preserve SG 31 \cite{ultrathin}, and should yield essentially the same effect of unconventional charge-to-spin conversion.

Our results for 2D-SnTe illustrate the power of a simple symmetry analysis for designing experimentally-relevant systems that exploit external control over the USHE. We note that similar materials could also potentially provide such control. For example, other compounds pertaining to SG 31, such as bulk 1T'-WTe2, may show a similar external control via an out-of-plane electric field or strain.


\section{Conclusions and outlook} \label{Conclusions}
In summary, we have determined spin Hall conductivity tensors for all 230 crystallographic space groups.
While the conventional spin Hall effect is universally present in all of them, the unconventional components are crystal symmetry selective.
We categorized spin Hall effects into conventional, collinear and longitudinal ones, analyzing important examples in each class, and providing a guide to design compounds suitable for applications.
Based on the lists of space groups allowing a specific type of spin Hall conductivity, further materials candidates can be easily found using crystal structures databases, such as AFLOWLIB or Materials Project \cite{aflowlib, materials_project}.

In addition, we have revealed that unconventional spin Hall components can be induced by an external electric field which breaks symmetries of certain crystals, leading to a change of the space group. We have verified this concept by performing DFT calculations for 2D-SnTe, and we have found that an additional component is as large as the conventional ones, suggesting the possibility of the experimental confirmation. Further systematic search could reveal several materials with similar or enhanced properties. We believe that the devices allowing spin injection tuned by the electric field will open exciting perspectives for spintronics.

Finally, we note that the simultaneous presence of several conventional and unconventional spin Hall components can be further explored towards the design of novel spintronic devices. One of the most interesting possibilities is to use materials with both collinear and collinear-longitudinal components, such as $\sigma^{y}_{yx}$ and $\sigma^{y}_{yy}$, in order to accomplish a more efficient switching mechanism of spin-orbit torques. Another option is a realization of elementary gates in spin logic circuits. We are convinced that these results and prospects will stimulate further research from the theoretical and experimental side.

\vskip 1.0 em
\begin{acknowledgments}
J.S. acknowledges Rosalind Franklin Fellowship from the University of Groningen. M.H.D.G. acknowledges the support from the Dutch Research Council (NWO - grant STU.019.014). The calculations were carried out on the Dutch national e-infrastructure with the support of SURF Cooperative, and on the Peregrine high performance computing cluster of the University of Groningen.
\end{acknowledgments}

%

\section*{Appendix}\label{appendix}
To obtain spin Hall conductivity tensors, we follow the convention from the Bilbao Crystallographic Server (BCS) and \textit{Physical Properties of Crystals} by John Nye (Appendix B) \cite{nye1985physical}. Calculation of any physical tensor using the TENSOR program requires a well defined orthogonal basis. An orthogonal basis ($\bm{a'},\bm{b'},\bm{c'}$) requires
\begin{equation}\label{ortho_lattice}
    \bm{a}'\; \|\; \bm{a},\;
    \bm{c}'\; \|\; \bm{c}^*,\;
    \bm{b}'\; \|\; \bm{c}'\times\bm{a}
\end{equation}
where ($\bm{a},\bm{b},\bm{c}$) are conventional crystal lattice vectors \cite{brock2016international} in a reference frame ($\bm{Ox} \| \bm{a}$, $\bm{Oy} \| \bm{b}$, $\bm{Oz} \| \bm{c}$). An orthogonal reference frame ($\bm{x}$, $\bm{y}$, $\bm{z}$) can be obtained similarly following Eq. (\ref{ortho_lattice}) which is required to obtain the symmetry allowed spin Hall conductivity tensors.

Below we list SHC tensors calculated for all 230 crystallographic space groups. Note that the form of the spin Hall conductivity tensor is determined by the 11 Laue classes, except for trigonal lattices, where the crystal class needs to be additionally specified.

\begin{table}[H]
\begin{tabular}{|m{4cm}|m{4cm}|}\hline
\begin{center}
\textbf{ \makecell{SG 1 - SG 2 \\ Laue class: -1} \vspace{-20pt}}\end{center}  &
\makecell{$\sigma^{x} = \begin{pmatrix}
     \sigma_{xx}^{x} & \sigma_{xy}^{x} & \sigma_{xz}^{x} \\
     \sigma_{yx}^{x} & \sigma_{yy}^{x} & \sigma_{yz}^{x} \\
     \sigma_{zx}^{x} & \sigma_{zy}^{x} & \sigma_{zz}^{x}
\end{pmatrix}$} \\ \hline
\centering 27 components &
\makecell{$\sigma^{y} = \begin{pmatrix}
     \sigma_{xx}^{y} & \sigma_{xy}^{y} & \sigma_{xz}^{y} \\
     \sigma_{yx}^{y} & \sigma_{yy}^{y} & \sigma_{xy}^{x} \\
     \sigma_{zx}^{y} & \sigma_{zy}^{y} & \sigma_{zz}^{y}
\end{pmatrix}$} \hrule \\
\centering 27 independent &
\makecell{$\sigma^{z} = \begin{pmatrix}
     \sigma_{xx}^{z} & \sigma_{xy}^{z} & \sigma_{xz}^{z} \\
     \sigma_{yx}^{z} & \sigma_{yy}^{z} & \sigma_{yz}^{z} \\
     \sigma_{zx}^{x} & \sigma_{zy}^{z} & \sigma_{zz}^{z}
\end{pmatrix}$}\\
\hline
\end{tabular}
\end{table}

\begin{table}[H]
\begin{tabular}{|m{4cm}|m{4cm}|}
\hline
\begin{center}
\textbf{ \makecell{SG 3 - SG 15 \\ Laue class: 2/m} \vspace{-20pt}}\end{center} &
\makecell{$\sigma^{x} = \begin{pmatrix}
     0 & \sigma_{xy}^{x} & 0 \\
     \sigma_{yx}^{x} & 0 & \sigma_{yz}^{x} \\
     0 & \sigma_{zy}^{x} & 0
\end{pmatrix}$} \\ \hline
\centering 13 components &
\makecell{$\sigma^{y} = \begin{pmatrix}
     \sigma_{xx}^{y} & 0 & \sigma_{xz}^{y} \\
     0 & \sigma_{yy}^{y} & 0 \\
     \sigma_{zx}^{y} & 0 & \sigma_{zz}^{y}
\end{pmatrix}$} \hrule \\
\centering 13 independent &
\makecell{$\sigma^{z} = \begin{pmatrix}
     0 & \sigma_{xy}^{z} & 0 \\
     \sigma_{yx}^{z} & 0 & \sigma_{yz}^{z} \\
     0 & \sigma_{zy}^{z} & 0
\end{pmatrix}$}\\
\hline
\end{tabular}
\end{table}

\vspace{-20pt}
\begin{table}[H]
\begin{tabular}{|m{4cm}|m{4cm}|}
\hline
\begin{center}
\textbf{ \makecell{SG 16 - SG 74 \\ Laue class: mmm} \vspace{-20pt}}\end{center}  &
\makecell{$\sigma^{x} = \begin{pmatrix}
     0 & 0 & 0 \\
     0 & 0 & \sigma_{yz}^{x} \\
     0 & \sigma_{zy}^{x} & 0
\end{pmatrix}$}\\ \hline
\centering 6 components &
\makecell{$\sigma^{y} = \begin{pmatrix}
     0 & 0 & \sigma_{xz}^{y} \\
     0 & 0 & 0 \\
     \sigma_{zx}^{y} & 0 & 0
\end{pmatrix}$} \\

\centering 6 independent & \hrule
\makecell{$\sigma^{z} = \begin{pmatrix}
     0 & \sigma_{xy}^{z} & 0 \\
     \sigma_{yx}^{z} & 0 & 0 \\
     0 & 0 & 0
\end{pmatrix}$}\\
\hline
\end{tabular}
\end{table}
\vspace{-20pt}
\begin{table}[H]
\begin{tabular}{|m{4cm}|m{4cm}|}
\hline
\begin{center}
\textbf{\makecell{SG 75 - SG 88 \\ Laue class: 4/m} \vspace{-20pt}}\end{center} &
\makecell{$\sigma^{x} = \begin{pmatrix}
     0 & 0 & \sigma_{xz}^{x} \\
     0 & 0 & \sigma_{yz}^{x} \\
     \sigma_{zx}^{x} & \sigma_{zy}^{x} & 0
\end{pmatrix}$}\\ \hline
\begin{center}
\makecell{13 components,\\ 7 independents\\
$\sigma_{zx}^{x} = \sigma_{zy}^{y}$,
$\sigma_{xz}^{x} = \sigma_{yz}^{y}$,}\end{center}&
\makecell{$\sigma^{y} = \begin{pmatrix}
     0 & 0 & \sigma_{xz}^{y} \\
     0 & 0 & \sigma_{yz}^{y} \\
     \sigma_{zx}^{y} & \sigma_{zy}^{y} & 0
\end{pmatrix}$} \hrule \\
\centering
$\sigma_{zx}^{y} = -\sigma_{zy}^{x}$,
$\sigma_{xz}^{y} = -\sigma_{yz}^{x}$,
$\sigma_{xx}^{z} = \sigma_{yy}^{z}$,
$\sigma_{xy}^{z} = -\sigma_{yx}^{z}$&
\makecell{$\sigma^{z} = \begin{pmatrix}
     \sigma_{xx}^{z} & \sigma_{xy}^{z} & 0 \\
     \sigma_{yx}^{z} & \sigma_{yy}^{z} & 0 \\
     0 & 0 & \sigma_{zz}^{z}
\end{pmatrix}$}\\
\hline
\end{tabular}
\end{table}
\vspace{-20pt}
\begin{table}[H]
\begin{tabular}{|m{4cm}|m{4cm}|}
\hline
\centering
\textbf{ \makecell{SG 89 - SG 142 \\ Laue class: 4/mmm} \vspace{-20pt}} &
\makecell{$\sigma^{x} = \begin{pmatrix}
     0 & 0 & 0 \\
     0 & 0 & \sigma_{yz}^{x} \\
     0 & \sigma_{zy}^{x} & 0
\end{pmatrix}$}\\
\hline \makecell{6 components, \\ 3 independents, \\
$\sigma_{xz}^{y} = -\sigma_{yz}^{x}$} &
\makecell{$\sigma^{y} = \begin{pmatrix}
     0 & 0 & \sigma_{xz}^{y} \\
     0 & 0 & 0 \\
     \sigma_{zx}^{y} & 0 & 0
\end{pmatrix}$}\\
\centering
\makecell{
$\sigma_{xy}^{z} = -\sigma_{yx}^{z}$,
$\sigma_{zx}^{y} = -\sigma_{zy}^{x}$ }&
\hrule \makecell{$\sigma^{z} = \begin{pmatrix}
     0 & \sigma_{xy}^{z} & 0 \\
    \sigma_{yx}^{z} & 0 & 0 \\
     0 & 0 & 0
\end{pmatrix}$}\\
\hline
\end{tabular}
\end{table}

\vspace{-20pt}
\begin{table}[H]
\begin{tabular}{|m{4cm}|m{4cm}|}
\hline
\begin{center}
\makecell{\textbf{SG 143 - SG 148} \\ \textbf{Laue class: -3}} \end{center} &
\makecell{$\sigma^{x} = \begin{pmatrix}
     \sigma_{xx}^{x} & \sigma_{xy}^{x} & \sigma_{xz}^{x} \\
     \sigma_{yx}^{x} & \sigma_{yy}^{x} & \sigma_{yz}^{x} \\
     \sigma_{zx}^{x} & \sigma_{zy}^{x} & 0
\end{pmatrix}$} \\\hline
\makecell{21 components,\\ 9 independent,\\
$\sigma_{xx}^{x}$ = - $\sigma_{yy}^{x} $ =
$-\sigma_{yx}^{y}$ =- $\sigma_{xy}^{y}$,\\
$\sigma_{zx}^{x}$ = $\sigma_{zx}^{y}$,} &
\makecell{$\sigma^{y} = \begin{pmatrix}
\sigma_{xx}^{y} & \sigma_{xy}^{y} & \sigma_{xz}^{y}\\
\sigma_{yx}^{y} & \sigma_{yy}^{y} & \sigma_{xy}^{x} \\
\sigma_{zx}^{y} & \sigma_{zy}^{y} & 0
\end{pmatrix}$} \\
\makecell{$\sigma_{xz}^{x} = \sigma_{yz}^{y}$,
$\sigma_{zx}^{y} = -\sigma_{zy}^{x}$,\\
$\sigma_{xx}^{y} = \sigma_{yx}^{x} =
-\sigma_{yy}^{y} = \sigma_{xy}^{x}$,\\
$\sigma_{xz}^{y} =  \sigma_{yz}^{x}$,
$\sigma_{xx}^{z} = \sigma_{yy}^{z}$,\\
$\sigma_{xy}^{z} = \sigma_{yy}^{z}$} &
\hrule\makecell{$\sigma^{z} = \begin{pmatrix}
\sigma_{xx}^{z} & \sigma_{xy}^{z} & 0 \\
\sigma_{yx}^{z} & \sigma_{yy}^{z} & 0 \\
0 & 0 & \sigma_{zz}^{z} \end{pmatrix}$}\\
\hline
\end{tabular}
\end{table}
\vspace{-20pt}
\begin{table}[H]
\begin{tabular}{|m{4cm}|m{4cm}|}
\hline
\textbf{ \makecell{
    SG 149, 151, 153, 
    157, \\
    159 162,163 \\
Laue class: -3m}} &
\makecell{$\sigma^{x} = \begin{pmatrix}
     0 & \sigma_{xy}^{x} & 0 \\
     \sigma_{yx}^{x} & 0 & \sigma_{yz}^{x} \\
     0 & \sigma_{zy}^{x} & 0
\end{pmatrix}$}\\ \hline
\centering \makecell{10 components,\\ 4 independent \\
$\sigma_{xx}^{y} = \sigma_{xy}^{x} =
\sigma_{yx}^{x} = -\sigma_{yy}^{y}$} &
\makecell{$\sigma^{y} = \begin{pmatrix}
     \sigma_{xx}^{y} & 0 & \sigma_{xz}^{y} \\
     0 & \sigma_{yy}^{y} & 0 \\
     \sigma_{zx}^{y} & 0 & 0
\end{pmatrix}$} \\
\makecell{
$\sigma_{zx}^{y} = -\sigma_{zy}^{x}$,
$\sigma_{xz}^{y} = -\sigma_{yz}^{x}$, \\
$\sigma_{xy}^{z} = - \sigma_{yx}^{z}$} &
\hrule
\makecell{$\sigma^{z} = \begin{pmatrix}
     0 & \sigma_{xy}^{z} & 0 \\
     \sigma_{yx}^{z} & 0 & 0 \\
     0 & 0 & 0
\end{pmatrix}$}\\
\hline
\end{tabular}
\end{table}
\vspace{-20pt}
\begin{table}[H]
\begin{tabular}{|m{4cm}|m{4cm}|}
\hline
\textbf{\makecell{SG 150, 152, 154-155, \\
156, 158, 160, 161, \\
164 - 167 \\
Laue class: -3m}} &
\makecell{$\sigma^{x} = \begin{pmatrix}
     \sigma_{xx}^{x} & 0 & 0 \\
     0 & \sigma_{yy}^{x} & \sigma_{yz}^{x} \\
     0 & \sigma_{zy}^{x} & 0
\end{pmatrix}$}\\ \hline
\makecell{ 10 components,\\ 4 independent \\
$\sigma_{xx}^{x}$ =$- \sigma_{yy}^{x}$ =
$-\sigma_{yx}^{y}$ = $-\sigma_{xy}^{y}$,} &
\makecell{$\sigma^{y} = \begin{pmatrix}
     0 & \sigma_{xy}^{y} & \sigma_{xz}^{y} \\
     \sigma_{yx}^{y} & 0 & 0 \\
     \sigma_{zx}^{y} & 0 & 0
\end{pmatrix}$} \hrule \\
\makecell{
$\sigma_{zx}^{y} = -\sigma_{zy}^{x}$,
$\sigma_{xz}^{y} = -\sigma_{yz}^{x}$,\\
$\sigma_{xy}^{z} = -\sigma_{yz}^{x}$} &
\makecell{$\sigma^{z} = \begin{pmatrix}
     0 & \sigma_{xy}^{z} & 0 \\
     \sigma_{yx}^{z} & 0 & 0 \\
     0 & 0 & 0
\end{pmatrix}$}\\
\hline
\end{tabular}
\end{table}
\vspace{-20pt}
\begin{table}[H]
\begin{tabular}{|m{4cm}|m{4cm}|}
\hline
\begin{center}
\textbf{ \makecell{SG 168 - SG 176 \\ Laue class: 6/m}\vspace{-20pt} }\end{center}  &
\makecell{$\sigma^{x} = \begin{pmatrix}
     0 & 0 & \sigma_{xz}^{x} \\
     0 & 0 & \sigma_{yz}^{x} \\
     \sigma_{zx}^{x} & \sigma_{zy}^{x} & 0
\end{pmatrix}$}\\\hline
\centering
\makecell{13 components,\\ 7 independent\\
$\sigma_{zx}^{x} = \sigma_{zy}^{y}$,
$\sigma_{xz}^{x} = \sigma_{yz}^{y}$,} &
\makecell{$\sigma^{y} = \begin{pmatrix}
     0 & 0 & \sigma_{xz}^{y} \\
     0 & 0 & \sigma_{yz}^{y} \\
     \sigma_{zx}^{y} & \sigma_{zy}^{y} & 0
\end{pmatrix}$} \\
\centering
\makecell{$\sigma_{zx}^{y} = -\sigma_{zy}^{x}$,
$\sigma_{xz}^{y} = -\sigma_{yz}^{x}$,\\
$\sigma_{xx}^{z} = \sigma_{yy}^{z}$,
$\sigma_{xy}^{z} = - \sigma_{yx}^{z}$}&\hrule \makecell{$\sigma^{z} = \begin{pmatrix}
     \sigma_{xx}^{z} & \sigma_{xy}^{z} & 0 \\
     \sigma_{yx}^{z} & \sigma_{yy}^{z} & 0 \\
     0 & 0 & \sigma_{zz}^{z}
\end{pmatrix}$}\\ \hline
\end{tabular}
\end{table}
\vspace{-25pt}
\begin{table}[H]
\begin{tabular}{|m{4cm}|m{4cm}|}
\hline
\centering
\textbf{ \makecell{SG 177 - SG 194 \\ Laue class: 6/mmm}\vspace{-20pt}} &
\makecell{$\sigma^{x} = \begin{pmatrix}
     0 & 0 & 0 \\
     0 & 0 & \sigma_{yz}^{x} \\
     0 & \sigma_{zy}^{x} & 0
\end{pmatrix}$}\\
\hline\centering 6 components, 3 independent &
\makecell{$\sigma^{y} = \begin{pmatrix}
     0 & 0 & 0 \\
     0 & 0 & \sigma_{xy}^{x} \\
     0 & \sigma_{zy}^{y} & 0
\end{pmatrix}$}  \\ \hline
\centering \makecell{$\sigma_{zx}^{y} = - \sigma_{zy}^{x}$,\\
$\sigma_{xz}^{y} = -\sigma_{yz}^{x}$,\\
$\sigma_{xy}^{z} = -\sigma_{yx}^{z}$} &
\makecell{$\sigma^{z} = \begin{pmatrix}
     0 & \sigma_{xy}^{z} & 0 \\
     \sigma_{yx}^{z} & 0 & 0 \\
     0 & 0 & 0
\end{pmatrix}$}\\
\hline
\end{tabular}
\end{table}
\vspace{-20pt}
\begin{table}[H]
\begin{tabular}{|m{4cm}|m{4cm}|}
\hline
\begin{center}
\textbf{ \makecell{SG 195 - SG 206 \\Laue class: m-3}\vspace{-20pt} }\end{center}  &
\makecell{$\sigma^{x} = \begin{pmatrix}
     0 & 0 & 0 \\
     0 & 0 & \sigma_{yz}^{x} \\
     0 & \sigma_{zy}^{x} & 0
\end{pmatrix}$}\\\hline
\centering 6 components, 2 independent &
\makecell{$\sigma^{y} = \begin{pmatrix}
     0 & 0 & \sigma_{xz}^{y} \\
     0 & 0 & 0 \\
     \sigma_{zx}^{y} & 0 & 0
\end{pmatrix}$} \hrule \\
\centering \makecell{$\sigma_{xy}^{z} = \sigma_{yz}^{x} = \sigma_{zx}^{y}$ \\
$\sigma_{xz}^{y} = \sigma_{zy}^{x} = \sigma_{yx}^{z}$} &
\makecell{$\sigma^{z} = \begin{pmatrix}
     0 & \sigma_{xy}^{z} & 0 \\
     \sigma_{yx}^{z} & 0 & 0 \\
     0 & 0 & 0
\end{pmatrix}$}\\
\hline
\end{tabular}
\end{table}
\vspace{-22pt}
\begin{table}[H]
\begin{tabular}{|m{4cm}|m{4cm}|}
\hline
\begin{center}
\textbf{ \makecell{SG 207 - SG 230 \\ Laue class: m-3m} \vspace{-20pt}}\end{center} &
\makecell{$\sigma^{x} = \begin{pmatrix}
     0 & 0 & 0 \\
     0 & 0 & \sigma_{yz}^{x} \\
     0 & \sigma_{zy}^{x} & 0
\end{pmatrix}$}\\ \hline
\centering 6 components, 1 independent &
\makecell{$\sigma^{y} = \begin{pmatrix}
     0 & 0 & \sigma_{xz}^{y} \\
     0 & 0 & 0 \\
     \sigma_{zx}^{y} & 0 & 0
\end{pmatrix}$} \hrule \\
\centering \makecell{$\sigma_{xy}^{z} = \sigma_{yz}^{x} = \sigma_{zx}^{y}=$\\
$- \sigma_{yx}^{z} = -\sigma_{zy}^{x} = -\sigma_{xz}^{y} $} &
\makecell{$\sigma^{z} = \begin{pmatrix}
     0   & \sigma_{xy}^{z} & 0 \\
     \sigma_{yx}^{z} & 0 & 0 \\
     0 & 0 & 0
\end{pmatrix}$}\\
\hline
\end{tabular}
\end{table}
\end{document}